\documentclass[aps,prd,preprint,balancelastpage,nofootinbib,preprintnumbers,amsmath,amssymb,superscriptaddress]{revtex4-1}
\usepackage{graphicx}   % include figures
\usepackage[
colorlinks=true,        % color link
citecolor=blue,         % cite color
linkcolor=blue,         % link color
urlcolor=blue           % url color
]{hyperref}  
\usepackage{tabulary}
\usepackage{color}      % \textcolor{declared-color}{text}
\usepackage{amsmath}    % 加载数学环境支持
\usepackage{amsfonts}   % 加载数学字体支持

\newcommand{\nc}{\newcommand*} 
\newcommand{\be}{\begin{equation}}
\newcommand{\ee}{\end{equation}}
\newcommand{\bea}{\setlength\arraycolsep{2pt} \begin{eqnarray}}
\newcommand{\eea}{\end{eqnarray}}

\nc{\al}{\alpha}
\nc{\s}{\sigma}
\nc{\dt}{\delta}
\nc{\Dt}{\Delta}
\nc{\Ld}{\Lambda}
\nc{\p}{\partial}
\nc{\om}{\omega}
\nc{\Om}{\Omega}
\nc{\rd}{\mathrm{d}}
\nc{\Od}[1]{\mathcal{O}(#1)} % order operator
\nc{\kp}{\kappa}

\def\[{\left[}
\def\]{\right]}
\def\e{\begin{equation}}
\def\q{\end{equation}}
\def\m{\begin{eqnarray}}
\def\n{\end{eqnarray}}

\nc{\Eq}[1]{Eq.~\eqref{#1}}     % equation
\nc{\Fig}[1]{Fig.~\ref{#1}}     % figure
\nc{\Table}[1]{Table~\ref{#1}}  % table
\nc{\Sec}[1]{Sec.~\ref{#1}}     % section

\nc{\Msun}{M_\odot}             % solar mass
\nc{\fpbhn}{f_{\mathrm{pbh0}}}    % f_pbh
\nc{\mR}{\mathcal{R}} % merger rate density
\nc{\seq}{\sigma_{\mathrm{eq}}}
\nc{\ogw}{\Omega_{\mathrm{GW}}}
\nc{\gpcyr}{\mathrm{Gpc}^{-3}\,\mathrm{yr}^{-1}}
\nc{\lvc}{LIGO/Virgo} % LIGO-VIRGO collaboration
\nc{\SNR}{\mathrm{SNR}} % signal to noise ratio
\nc{\mmin}{{m_{\mathrm{min}}}}
\nc{\mmax}{{m_{\mathrm{max}}}}
\nc{\Mmin}{{M_{\mathrm{min}}}}
\nc{\fmin}{{f_{\mathrm{min}}}}
\nc{\VT}{\mathrm{VT}}
\nc{\rhoGW}{\rho_{\mathrm{GW}}}
\nc{\vth}{\vec{\theta}}
\nc{\vd}{\vec{d}}
\nc{\vla}{\vec{\lambda}}
\nc{\Nobs}{N_{\mathrm{obs}}}
\nc{\av}[1]{\langle #1 \rangle} % average bracket
\nc{\km}{\mathrm{km}}
\nc{\Mpc}{\mathrm{Mpc}}
\nc{\Tobs}{T_{\mathrm{obs}}}
\nc{\Ntemp}{N_{\mathrm{temp}}}
\nc{\ie}{\textit{i.e.}}
\nc{\eg}{\textit{e.g.~}}
\nc{\app}{\approx}
\nc{\hf}{\frac{1}{2}}

\nc{\mpbh}{m_{\rm{pbh}}}
\nc{\cR}{\mathcal{R}}
\nc{\mU}{{\mathcal{U}}}
\nc{\Mc}{{M_\mathrm{c}}}
\nc{\Mf}{{M_\mathrm{f}}}
\nc{\red}[1]{\textcolor{red}{#1}}
\nc{\yellow}[1]{\textcolor{yellow}{#1}}
\nc{\green}[1]{\textcolor{green}{#1}}
\nc{\blue}[1]{\textcolor{blue}{#1}}

\begin{document}
%\title{\textcolor{red}{Probing the Microphysical Reheating History of Decaying Oscillatory Inflation with ACT and Planck Constraints}}

\title{Probing the Perturbative Reheating History of Decaying Oscillatory Inflation with ACT Constraints}

%%%%%%%%%%%%%%%%%%%%%%%%%%%%%%%%%%%% author %%%%%%%%%%%%%%%%%%%%%%%%%%%%%%%%%%%%
\author{Li-Yang Chen}
\email{lychen@cdnu.edu.cn}
\affiliation{College of Physics and Electronic Engineering, Chengdu Normal University, Chengdu, Sichuan, 611130, China}
%\affiliation{Institute of Interdisciplinary Studies, Hunan Normal University, Changsha, Hunan 410081, China}

%%%%%%%%%%%%%%%%%%%%%%%%%%%%%%%%%%%% author %%%%%%%%%%%%%%%%%%%%%%%%%%%%%%%%%%%%
\author{Rongrong Zhai}
\email{rrzhai@xztu.edu.cn}
\affiliation{Department of Physics, Xinzhou Normal University, Xinzhou 034000, Shanxi, China}

%%%%%%%%%%%%%%%%%%%%%%%%%%%%%%%%%%%% author %%%%%%%%%%%%%%%%%%%%%%%%%%%%%%%%%%%%
\author{Feng-Yi Zhang}
\email{zfy@usc.edu.cn} % 通讯作者邮箱
\thanks{Corresponding author}
\affiliation{School of Mathematics and Physics, University of South China, Hengyang, 421001, China}

%%%%%%%%%%%%%%%%%%%%%%%%%%%%%%%%%%%% author %%%%%%%%%%%%%%%%%%%%%%%%%%%%%%%%%%%%

\begin{abstract}

Precision measurements of the Cosmic Microwave Background (CMB) now offer a powerful probe of the unknown reheating epoch. In this work, we scrutinize a decaying oscillatory inflation model inspired by supergravity, replacing standard ad hoc reheating assumptions with a fully dynamical calculation based on perturbative inflaton decay. By numerically tracking the energy transfer and the evolution of the equation of state, we eliminate the theoretical degeneracy associated with the reheating duration, directly linking the microphysical decay rate $\Gamma$ to the observable spectral index $n_s$. We confront these self-consistent predictions with the combined constraints from Planck 2018 and ACT DR6.  Our analysis demonstrates that the viable parameter space is tightly bracketed: the thermalization requirement from Big Bang Nucleosynthesis imposes a strict lower bound on the coupling strength, while the latest ACT data strongly favor scenarios with efficient reheating ($T_{\text{re}} \gtrsim 10^{14}$ GeV), effectively pushing the model towards the instantaneous reheating limit. This study highlights the capability of modern CMB data to constrain the particle physics nature of the early universe.

\end{abstract}

\maketitle

%%%%%%%%%%%%%%%%%%%%%%%%%%%%%%%%%%%%%%%%%%%%%%%%%%%%%%%%%%%%%%%%%

\section{Introduction}
\label{sec:introduction}

Cosmic inflation offers compelling solutions to several fundamental problems of the early universe and is strongly supported by observations of the cosmic microwave background (CMB), which provide precise constraints on the scalar spectral index $n_s$ and the tensor-to-scalar ratio $r$~\cite{STAROBINSKY198099, PhysRevLett.48.1220, PhysRevD.23.347, LINDE1982389}. High-precision measurements from Planck 2018 constrain $n_s = 0.9651\pm 0.0044$~\cite{akrami2020planck}, while the BICEP/Keck 2018 (BK18) results place a stringent upper bound on the tensor amplitude, $r_{0.05} < 0.036$ at 95\% confidence level~\cite{PhysRevLett.127.151301}. Recently, observations from the Atacama Cosmology Telescope (ACT) Collaboration \cite{louis2025atacamacosmologytelescopedr6,calabrese2025atacamacosmologytelescopedr6} reported a modest upward shift in the inferred value of $n_s$ relative to Planck 2018. When combined with Planck data (P-ACT), the scalar spectral index is determined to be $n_s = 0.9709 \pm 0.0038$, indicating a $\sim 2\sigma$ deviation from the Planck-only result. Incorporating additional datasets, such as CMB lensing and baryon acoustic oscillation (BAO) measurements from the Dark Energy Spectroscopic Instrument (DESI)~\cite{Adame_2025, Adame2_2025}, refines this constraint further. The joint analysis of Planck, ACT, and LiteBIRD (P-ACT-LB) yields $n_s = 0.9743 \pm 0.0034$, representing a statistically significant deviation and thus a tighter constraint on viable inflationary scenarios. A broad class of inflationary models---including hilltop, attractor, polynomial, plateau, and chaotic types---have therefore been revisited in light of the ACT data, often leading to stronger exclusions or increased parameter tuning~\cite{lynker2025actimplicationshilltopinflation,pallis2025actinspiredkaehlerbasedinflationaryattractors, peng2025polynomialpotentialinflationlight,haque2025minimalplateauinflationlight, Pallis_2025,wolf2025inflationaryattractorsradiativecorrections}. For completeness, we note that current constraints on small-scale curvature perturbations remain relatively weak, leaving open the possibility of enhanced perturbations and their associated physical implications~\cite{Pi_2018,Wang_2024,Chen:2022dqr,Chen:2024gqn,wang2025primordialblackholesscalar,2023PhRvD.108d3529Z,2022PhRvD.106b3517Z,Fu_2023,Fu_20231,Peng_2021,zhang2025gravitationalwavesgaugequanta,PPTA:2024xbe,murata2025stochastictailcurvatureperturbationhybrid,wang2025primordialblackholessave,chen2025primordialblackholesanisotropic,huang2025constantrollinflationprimordialblack}.

Beyond the inflationary phase itself, the subsequent reheating epoch plays a pivotal role in shaping the thermal history of the universe. By transferring the inflaton energy into relativistic degrees of freedom, reheating determines how the Universe enters the radiation-dominated era and sets the thermal conditions for the standard hot Big-Bang cosmology. Importantly, the post-inflationary expansion history feeds back into inflationary predictions through the relation between the pivot-scale $e$-fold number $N_*$ and the subsequent thermal evolution, thereby affecting the inferred values of $n_s$ and $r$~\cite{PhysRevD.82.023511, PhysRevLett.114.081303, Cook_2015, PhysRevD.92.063506, PhysRevD.103.103540,PhysRevLett.113.041302, PhysRevD.104.103530, PhysRevD.93.083524, ZHANG2023101169, PhysRevD.95.103502,Goswami_2018, Maity_2019, DENG2022101135,Zhang:2025aak, Zhang:2025lfx,Zhang:2025tpg, Mishra_2021,Gong_2015, PhysRevD.102.021301, Garcia:2023tkk, Germ_n_2023,ZHANG2024101482,ZHANG2024138765,odintsov2025actinflationinfluencereheating}.
In these works it is modeled phenomenologically by a constant effective equation-of-state $w_{\rm re}$ and a free duration $N_{\rm re}$, which is convenient but can obscure the underlying microphysics and time-dependent dynamics.
Moreover, observational constraints from Big Bang Nucleosynthesis (BBN) impose robust lower bounds on the reheating temperature, further restricting the viable parameter space ~\cite{Kawasaki_1999, Kawasaki_2000, Hannestad_2004, Hasegawa_2019, Kawasaki:1999na, Hasegawa:2019jsa}. The enhanced precision of ACT data not only sharpens constraints on inflationary dynamics but also motivates a revisitation of reheating scenarios in concrete models. In particular, recent studies show that ACT DR6 data can already place nontrivial bounds on reheating parameters ~\cite{Drees:2025ngb, Zharov:2025evb}. Relatedly, reconciling the Planck-favored Starobinsky model with ACT observations may require an extended inflationary duration and a lower reheating temperature~\cite{risdianto2025preheatingstagestarobinskyinflation}, while nonminimally coupled Higgs inflation can be brought into agreement with ACT and BBN data under reheating assumptions corresponding to a relatively stiff effective equation of state ~\cite{liu2025reconcilinghiggsinflationact}. These examples highlight that precision CMB measurements can constrain not only the inflationary potential but also the post-inflationary thermal history.

Motivated by these observational advances and the resulting theoretical challenges, in this work we investigate a class of decaying oscillatory inflationary models inspired by minimal no-scale supergravity~\cite{chen2025model}. The scalar potential features a quasi-flat plateau at large field values that supports slow-roll inflation and transitions to a damped oscillatory form near the minimum, thereby naturally ending inflation and initiating reheating. We confront the predicted CMB observables with the latest ACT-based constraints, highlighting how the oscillatory structure impacts the model trajectories in the $(n_s,r)$ plane. To connect inflation to the hot Big-Bang initial conditions in a physically controlled and self-consistent manner, we adopt a dynamical reheating description based on perturbative inflaton decay into a radiation bath. By numerically solving the coupled inflaton--radiation system, we determine the reheating duration, the reheating temperature, and an effective equation of state directly from the time evolution, and combine them with the standard thermodynamic matching relation to obtain a consistent mapping between the post-inflationary thermal history and the pivot-scale inflationary predictions.

This paper is organized as follows. In Sec.~\ref{sec:inflation_dynamics} we introduce the model and present the inflationary dynamics and predictions. In Sec.~\ref{sec:reheating} we describe the dynamical reheating framework and perform the thermodynamic matching to the CMB pivot scale. We conclude in Sec.~\ref{sec:conclusions}. Throughout, we use the metric signature $(-,+,+,+)$ and set $M_{\mathrm{pl}}=1$.

\section{Inflationary Predictions}
\label{sec:inflation_dynamics}

We consider inflation driven by a single, minimally coupled scalar field, governed by the action
\begin{align}
\mathcal{S} = \int d^4 x \sqrt{-g} \left[\frac{1}{2}R - \frac{1}{2}g^{\mu\nu}\partial_\mu \phi \partial_\nu \phi - V(\phi) \right],
\end{align}
where $g$ is the determinant of the spacetime metric and $R$ is the Ricci scalar, and $V(\phi)$ denotes the scalar potential.
In this work, we investigate a class of inflationary models characterized by a decaying oscillatory potential~\cite{chen2025model}, given by
\begin{align}  \label{VV}
V(\phi) = \lambda\, \phi^{2n}\, \sin^2\left(\frac{l}{\phi^n}\right),
\end{align}
where \( n \) and \( l \) are dimensionless parameters that characterize the shape of the potential. 
As discussed in Ref.~\cite{chen2025model}, the potential features a flat plateau at large field values—suitable for supporting slow-roll inflation—and exhibits damped oscillations near the minimum, which naturally terminate inflation and trigger the onset of reheating.
The parameter \( \lambda \) is determined by matching the amplitude of the primordial curvature power spectrum to observational data. According to the 2018 Planck results~\cite{akrami2020planck}, this amplitude is given by
\begin{align}
\ln(10^{10} A_s) = 3.044 \pm 0.014.
\label{eq:logAs}
\end{align}
For completeness, we briefly outline the no-scale supergravity motivation for the effective potential adopted in Eq.~(\ref{VV}).
This oscillatory single-field potential can be viewed as being inspired by a minimal no-scale SUGRA sector based on the coset
$SU(2,1)/(SU(2)\times U(1))$, in which the scalar sector typically contains a matter-like chiral field $\Phi$ and a modulus $T$.
Starting from a minimal no-scale K\"ahler potential and a renormalizable Wess--Zumino superpotential, and assuming that the modulus is effectively stabilized and can be approximately fixed (e.g.\ $T+T^*\equiv c$), the inflaton-sector potential can be written in the standard no-scale form $V(\Phi)$.
Furthermore, by restricting the dynamics to an appropriate inflationary trajectory in field space (for instance, by treating $\Phi$ as an effective function of the canonically normalized inflaton $\phi$, such as
$\Phi(\phi)=A\,\phi^{n}\sin\!\left(l/\phi^{n}\right)$),
one may obtain, within this effective description, a single-field potential $V(\phi)$ featuring an oscillatory modulation.
In the parameter regime relevant for CMB-compatible solutions, where $|\Phi|^2\ll 3c$ and the leading terms dominate along the trajectory, $V(\phi)$ can be approximately reduced to an effective oscillatory form with the same functional dependence as Eq.~(\ref{VV}).

The background dynamics of the inflaton field are governed by the Friedmann and Klein-Gordon equations:
\begin{align}
3H^{2} &= \frac{1}{2} \dot{\phi}^{2}+V(\phi), \label{Fe} \\
\ddot{\phi}  + &3 H \dot{\phi} +V^{\prime}\left(\phi\right) = 0, \label{Fq}
\end{align}
where $H \equiv \dot{a}/a$ is the Hubble parameter, the overdot denotes derivatives with respect to cosmic time $t$, while the prime denotes derivatives with respect to the scalar field $\phi$.
To analyze the inflationary dynamics under the slow-roll approximation, we define the slow-roll parameters as
\begin{align}
\epsilon_V \equiv \frac{1}{2}\left(\frac{V'}{V}\right)^2, \qquad
\eta_V \equiv \frac{V''}{V}.
\end{align}
The number of $e$-folds between the horizon exit of the CMB pivot scale and the end of inflation is given by
\begin{align}
N_* &\approx \int_{\phi_*}^{\phi_{\mathrm{end}}} \frac{V}{V'}\, d\phi \notag \\
    &= -\frac{1}{2n} \int_{\phi_*}^{\phi_{\mathrm{end}}} \frac{\phi^{1+n}}{\phi^n - l \cot(l\phi^{-n})}\, d\phi, \label{Ns}
\end{align}
where $\phi_*$ is the field value at the horizon exit of the pivot scale, and $\phi_{\text{end}}$ denotes the field value at the end of inflation, determined by the condition $\epsilon_V(\phi_{\text{end}}) = 1$ or $\eta_V(\phi_{\text{end}}) = 1$.
Under the slow-roll approximation, the inflationary observables are given by
\begin{align}
n_s &\simeq 1 - 6\epsilon_V + 2\eta_V \notag \\
    &= \phi_*^{-2(1+n)} \bigg\{
      \left[-4n(1+n) + \phi_*^{2} \right] \phi_*^{2n}
      + 4nl(1+3n)\phi_*^{n} \cot\left(l \phi_*^{-n}\right) \notag \\
    &\quad\quad - 4n^{2}l^{2} \left[1 + 2\cot^{2} \left(l \phi_*^{-n}\right)\right]
    \bigg\}, \label{eq:ns} \\
r &\simeq 16\epsilon_V 
  = 32n^2\, \phi_*^{-2(1+n)} \left[\phi_*^n - l \cot\left(l\phi_*^{-n}\right)\right]^2. \label{eq:r}
\end{align}

\begin{figure}[htbp]
    \centering
    \includegraphics[width=0.55\textwidth]{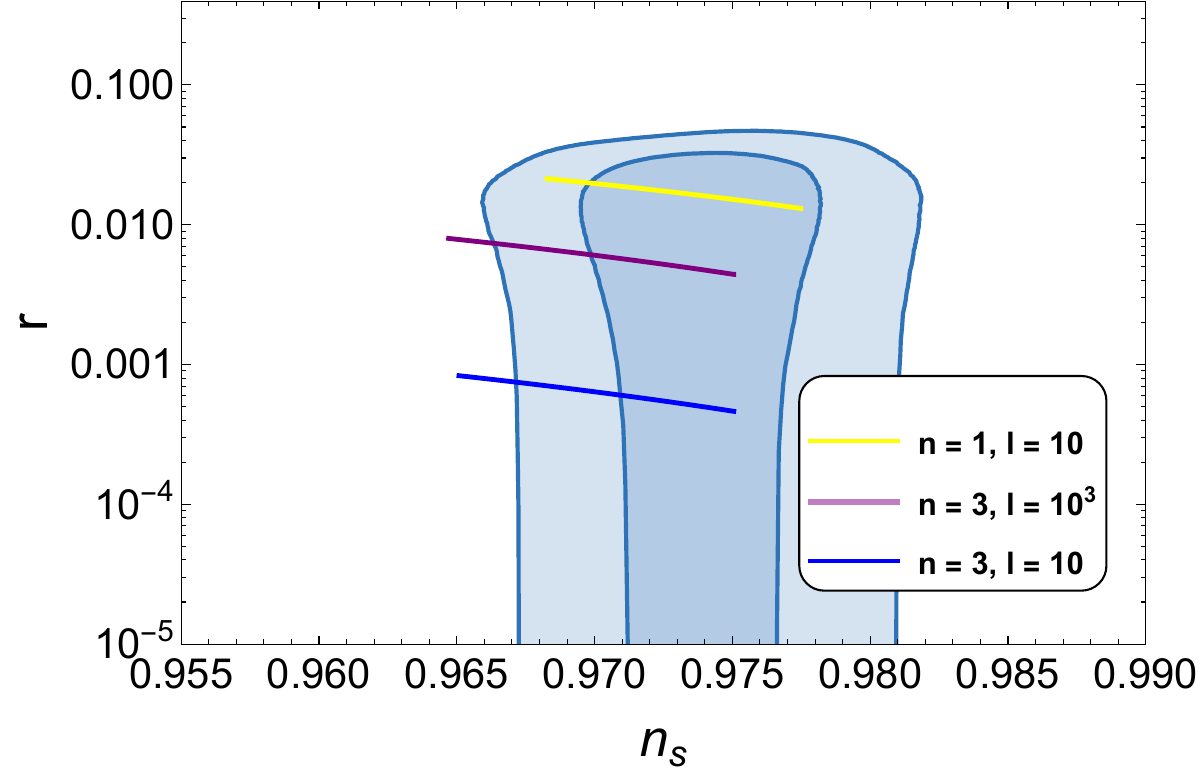}
    \caption{Predictions for the tensor-to-scalar ratio $r$ versus the scalar spectral index $n_s$
    in the decaying oscillatory inflationary model. The three solid curves correspond to the
    representative parameter choices $(n,l)=(1,10)$ (yellow), and $(3,10)$ (blue), $(3,10^{3})$ (purple).
    Along each curve, the pivot-scale $e$-fold number is varied in the range $N_*\in[50,70]$.
    The dark and light shaded regions denote the $1\sigma$ and $2\sigma$ confidence intervals from
    the combined P-ACT-LB-BK18 data~\cite{Adame_2025,Adame2_2025}.}
    \label{fig:cmb}
\end{figure}

In Fig.~\ref{fig:cmb}, we present the predictions of the model in the $(n_s,r)$ plane and focus on three representative parameter choices for a detailed comparison. We emphasize that this selection is not arbitrary: it follows the benchmark parameter sets previously discussed in Ref.~\cite{chen2025model}, which provides a useful reference for comparison and allows us to clearly illustrate how the potential parameters affect the observable quantities.
Concretely, we consider three discrete representative points, $(n,l)=(1,10)$, $(3,10)$, and $(3,10^{3})$, and isolate the role of each parameter through pairwise comparisons. On the one hand, comparing $(1,10)$ and $(3,10)$ (with $l=10$ fixed) highlights the impact of the power-law index $n$. On the other hand, comparing $(3,10)$ and $(3,10^{3})$ (with $n=3$ fixed) highlights the effect of the oscillation-scale parameter $l$.
As shown in the figure, for fixed $l=10$, changing $n$ from $1$ to $3$ significantly suppresses the tensor-to-scalar ratio $r$, while $n_s$ also exhibits a noticeable leftward shift.
For fixed $n=3$, changing $l$ from $10$ to $10^{3}$ shifts the trajectory upward, enhancing $r$ with only a modest shift in $n_s$. Overall, scanning $N_*\in[50,70]$ generates trajectories that lie within (or pass through) the observationally allowed region from the current P-ACT-LB-BK18 data, demonstrating that these representative parameter choices are compatible with the latest observational constraints.

\section{Reheating Dynamics}
\label{sec:reheating}

Following inflation, the Universe must undergo reheating, during which the inflaton energy is converted into relativistic particles, initiating radiation domination and recovering the standard hot Big-Bang cosmology~\cite{Kofman:1997yn,Shtanov:1995ce,Davidson_2000,Kofman:1994rk}. Since the reheating duration and energy-transfer efficiency feed into the thermodynamic matching relation and thereby affect the pivot-scale $e$-fold number $N_*$, a physically controlled reheating description is essential for reliably connecting inflationary observables $(n_s,r)$ to the underlying parameter space. In this work we adopt perturbative reheating as a minimal dynamical mechanism: we model the energy injection into a radiation bath by a constant decay rate $\Gamma$ and numerically solve the coupled evolution equations for the inflaton and radiation energy densities, which self-consistently yields the time-dependent equation of state, the end of reheating, and the reheating temperature. We now formulate this perturbative-decay setup and present the background equations and quantities used to characterize the reheating stage.

We model perturbative reheating by assuming that the inflaton field $\phi$ decays into a radiation bath with
energy density $\rho_r$ at a constant decay rate $\Gamma$. The background dynamics during reheating is governed by the coupled
Boltzmann--Klein--Gordon system
\begin{align}
\ddot{\phi} + (3H + \Gamma)\dot{\phi} + V'(\phi) = 0,
\label{eq:KG_reheating} \\
\dot{\rho}_r + 4H\rho_r = \Gamma \dot{\phi}^{2},
\label{eq:radiation_reheating} \\
3H^2 = \frac{1}{2}\dot{\phi}^2 + V(\phi) + \rho_r,
\label{eq:Friedmann_reheating}
\end{align}
where the radiation component is treated as a relativistic fluid with $p_r=\rho_r/3$, which effectively assumes
rapid thermalization of the decay products.

For convenience, we measure time by the number of
$e$-folds after the end of inflation,
\begin{equation}
N \equiv \ln\!\left(\frac{a}{a_{\rm end}}\right),
\label{eq:Ndef}
\end{equation}
where $a_{\rm end}$ is the scale factor at the end of inflation. To characterize the macroscopic reheating evolution
and quantify the approach to radiation domination, we monitor the total equation-of-state (EoS) parameter. We further
introduce the inflaton energy density and pressure, $\rho_\phi \equiv \dot{\phi}^{2}/2 + V(\phi)$ and
$p_\phi \equiv \dot{\phi}^{2}/2 - V(\phi)$, and define
\begin{align}
w(N) \equiv \frac{p_\phi+p_r}{\rho_\phi+\rho_r}
= \frac{\frac{1}{2}\dot{\phi}^{2}-V(\phi)+\rho_r/3}{\frac{1}{2}\dot{\phi}^{2}+V(\phi)+\rho_r}.
\label{eq:w_def}
\end{align}
To facilitate a direct comparison with the usual constant-$w_{\rm re}$
parametrization, we introduce the cumulative $e$-fold average for $N>0$,
\begin{align}
\bar{w}(N) \equiv \frac{1}{N}\int_{0}^{N} w(\tilde N)\,d\tilde N.
\label{eq:wbar_def}
\end{align}
Here and in the following, the subscript ``re'' denotes quantities evaluated at the end of reheating, i.e.,
at $t=t_{\rm re}$ (or equivalently $N=N_{\rm re}$); in particular, we define the effective reheating EoS as
$w_{\rm re}\equiv \bar{w}(N_{\rm re})$.

The numerical integration is initialized at the end of inflation, with the initial conditions
$(\phi_{\rm end},\dot{\phi}_{\rm end},H_{\rm end})$ inherited from the inflationary solution in
Sec.~\ref{sec:inflation_dynamics}. We assume that the radiation component is negligible at $t_{\rm end}$ and set
$\rho_r(t_{\rm end})=0$. We then evolve Eqs.~\eqref{eq:KG_reheating}--\eqref{eq:Friedmann_reheating} forward in time
and define the end of reheating operationally by the equality of the inflaton and radiation energy densities,
\begin{align}
\rho_r(t_{\rm re}) = \rho_\phi(t_{\rm re}),
\label{eq:reheating_end}
\end{align}
which marks the transition to radiation domination and fixes the reheating duration as
$N_{\rm re}\equiv \ln(a_{\rm re}/a_{\rm end})=N(t_{\rm re})$.
In what follows, we illustrate the reheating dynamics for the same three representative parameter choices used in Fig.~\ref{fig:cmb}, namely $(n,l)=(1,10)$, $(3,10)$, and $(3,10^3)$, and adopt $\Gamma=10^{-6}$ as our baseline. 
This choice comfortably satisfies the BBN requirement. In the perturbative reheating picture, radiation domination is reached when $H\simeq\Gamma$, which implies
$
T_{\rm re}\sim \left({90}/{\pi^2 g_\ast}\right)^{1/4}\sqrt{\Gamma }\, ,
$
so the condition $T_{\rm re}\gtrsim \mathcal{O}({\rm MeV})$ only translates into a very weak lower bound on $\Gamma$, far below our fiducial value. 
Numerically, $\Gamma \simeq 2.435\times 10^{12}\,{\rm GeV}$
ensures efficient energy transfer into radiation within our integration window and a rapid approach to radiation domination. 
We have also tested smaller decay rates, but lowering $\Gamma$ by about two orders of magnitude makes reheating difficult to complete within the same time span, with $\rho_r$ typically remaining subdominant and not overtaking $\rho_\phi$. 
We therefore take $\Gamma \simeq 2.435\times 10^{12}\,{\rm GeV}$ as our fiducial choice in the benchmark analysis.

\begin{figure}[htbp]
    \centering
    \includegraphics[width=0.32\textwidth]{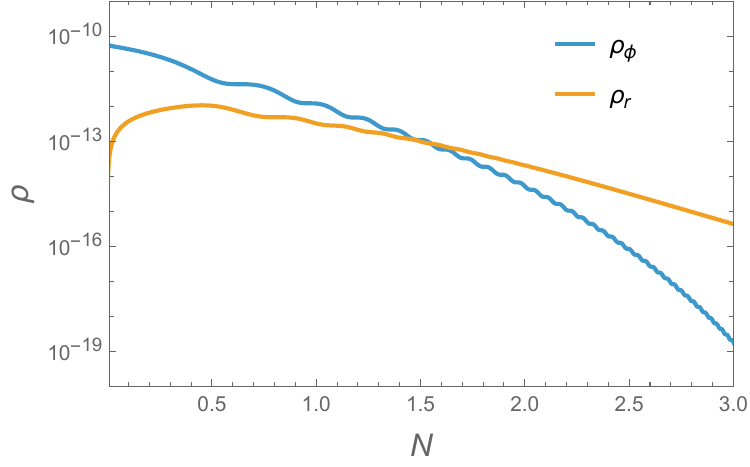}\hfill
    \includegraphics[width=0.32\textwidth]{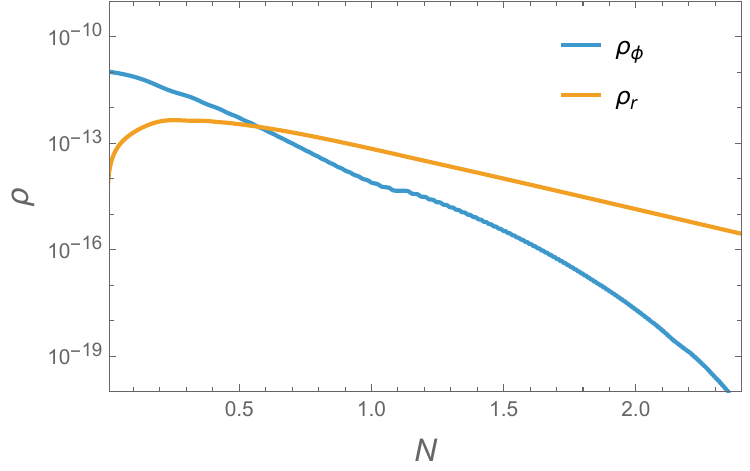}\hfill
    \includegraphics[width=0.32\textwidth]{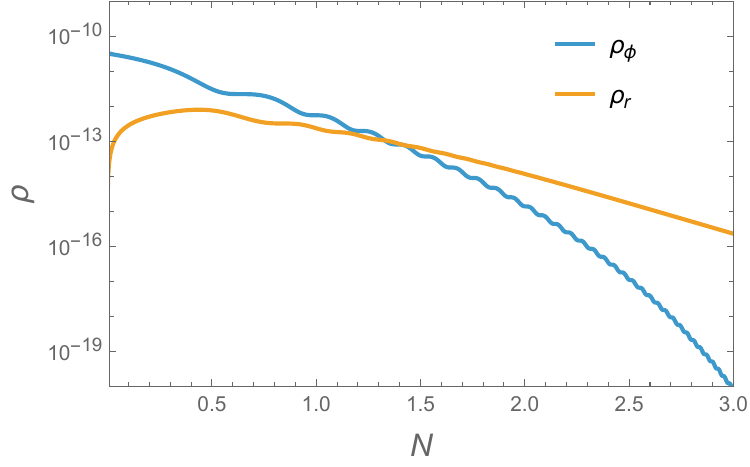}
    \caption{Numerical evolution of $\rho_\phi$ (blue) and $\rho_r$ (orange) as functions of $N$ during reheating for $\Gamma \simeq 2.435\times 10^{12}\,{\rm GeV}$, shown for three representative parameter choices. From left to right: $(n,l)=(1,10)$, $(3,10)$, and $(3,10^3)$.} 
    \label{fig:energy_density}
\end{figure}

Fig.~\ref{fig:energy_density} shows the numerical evolution of the inflaton energy density $\rho_\phi$ and the radiation energy density $\rho_r$ as functions of $N\equiv\ln(a/a_{\rm end})$. As the inflaton undergoes coherent oscillations around the minimum of the potential and injects energy into radiation through the decay term, $\rho_\phi$ decreases rapidly due to cosmic expansion, while $\rho_r$ gradually accumulates, driven by the source term $\Gamma\dot{\phi}^{\,2}$, and eventually crosses $\rho_\phi$. We operationally define the reheating completion time $t_{\rm re}$ by the equality $\rho_r=\rho_\phi$, which determines the reheating duration in $e$-folds as $N_{\rm re}=N(t_{\rm re})$. The crossing occurs at different values of
$N$ for different parameter choices, reflecting variations in the efficiency of energy transfer and the background evolution; nevertheless, all cases exhibit the characteristic transition from inflaton domination to radiation domination.

\begin{figure}[htbp]
    \centering
    \includegraphics[width=0.32\textwidth]{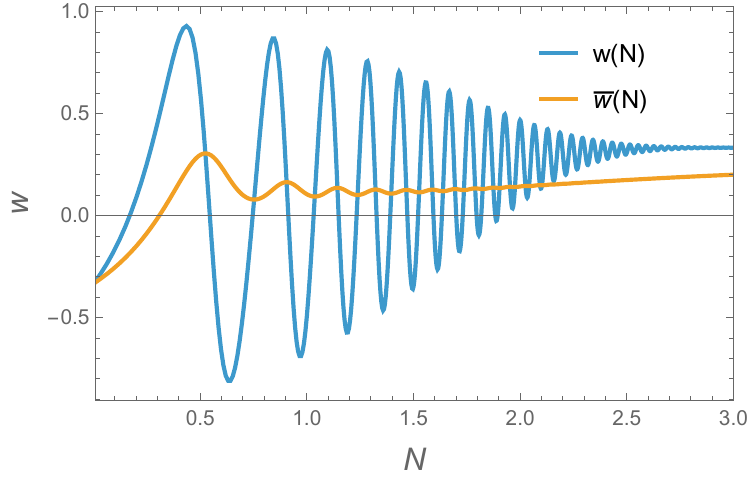}\hfill
    \includegraphics[width=0.32\textwidth]{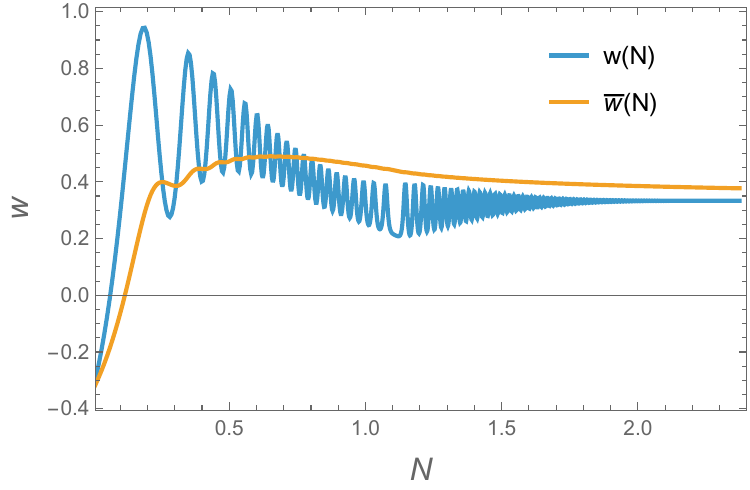}\hfill
    \includegraphics[width=0.32\textwidth]{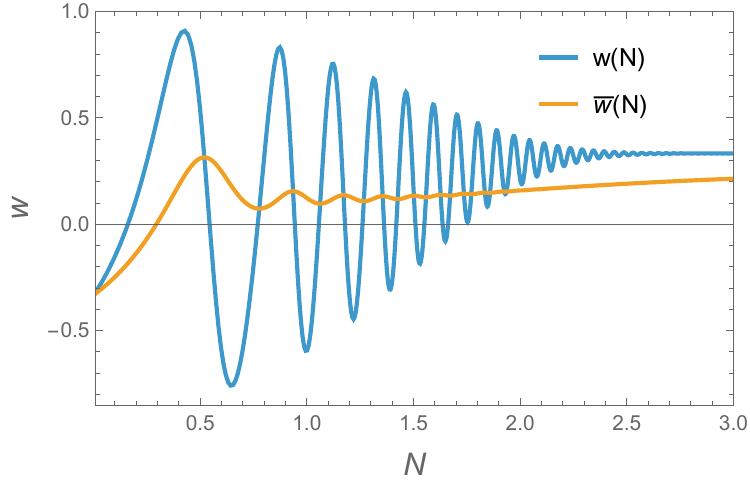}
    \caption{Evolution of the instantaneous equation-of-state parameter $w(N)$ (blue) and its cumulative average $\bar w(N)$ (orange) during reheating for $\Gamma \simeq 2.435\times 10^{12}\,{\rm GeV}$, shown for three representative parameter choices. From left to right: $(n,l)=(1,10)$, $(3,10)$, and $(3,10^3)$.} 
    \label{fig:eos_evolution}
\end{figure}

Fig.~\ref{fig:eos_evolution} displays the evolution of the instantaneous equation-of-state parameter $w(N)$ and its cumulative
average $\bar{w}(N)$. The rapid oscillations of $w(N)$ originate from coherent inflaton oscillations and decay in amplitude with
time, while the envelope gradually approaches $1/3$, signaling the approach to a radiation-dominated Universe. To characterize
the macroscopic expansion history, we also examine the cumulative average $\bar{w}(N)$ (orange curve), which is seen to remain
relatively flat (quasi-constant) over a sizable portion of reheating before drifting toward $1/3$. 

A notable qualitative difference arises in the middle panel compared to the left and right panels. For each set of model
parameters there exists a critical value $l_c$ that separates two distinct post-inflationary behaviors \cite{chen2025model}. When $l>l_c$, the inflaton exits inflation with a relatively small velocity and cannot roll through the first hill of the potential; instead, it becomes trapped
and oscillates around the first minimum $\phi_i$ satisfying $V(\phi_i)=0$. This regime (left and right panels) produces the familiar
alternating (sign-changing) oscillations of $w(N)$, while $\bar{w}(N)$ stays nearly constant for most of the reheating stage and then
relaxes to $1/3$ as radiation takes over. In contrast, when $l<l_c$ the inflaton is able to pass the first hill and keeps rolling across
the oscillatory structure before settling. This leads to a transient phase in which $w(N)$ remains predominantly positive and the
cumulative average $\bar{w}(N)$ can temporarily take a larger value (as seen in the middle panel), before eventually approaching
$1/3$. Therefore, the distinct behavior of the middle trajectory is a direct consequence of being in the $l<l_c$ branch, whereas the
left and right cases lie in the $l>l_c$ branch.

Once $t_{\rm re}$  is determined by Eq.~\eqref{eq:reheating_end}, the reheating temperature is obtained
from the radiation energy density at that moment,
\begin{align}
T_{\rm re} = \left(\frac{30\,\rho_r(t_{\rm re})}{\pi^2 g_{\rm re}}\right)^{1/4},
\label{eq:Tre_def}
\end{align}
where we take $g_{\rm re}\simeq 106.75$.
To connect the reheating outcome $(N_{\rm re},T_{\rm re})$ to the inflationary prediction at the CMB pivot scale
$k_*=0.05\,{\rm Mpc}^{-1}$, we use the standard thermodynamic matching relation~\cite{Cook_2015},
\begin{align}
N_* + N_{\rm re} + \ln\!\left(\frac{T_{\rm re}}{H_*}\right)
\simeq \ln\!\left[\frac{a_0T_\gamma}{k_*}\left(\frac{43}{11g_{\rm re}}\right)^{1/3}\right]
\simeq 60.8,
\label{eq:matching}
\end{align}
where $T_\gamma=2.7255\,{\rm K}$ and $a_0=1$. In our setup, both $H_*$ and $N_*$ are functions of $\phi_*$ determined by the inflationary dynamics
in Sec.~\ref{sec:inflation_dynamics}. Given the reheating outcome $(N_{\rm re},T_{\rm re})$ from the dynamical evolution,
Eq.~\eqref{eq:matching} fixes the consistent value of $\phi_*$, and hence determines $N_*$ uniquely. The corresponding CMB
observables $(n_s,r)$ are then evaluated using the slow-roll expressions in Eqs.~\eqref{eq:ns} and~\eqref{eq:r}.
In this way, the reheating dynamics closes the loop between the post-inflationary thermal history and the CMB-scale
inflationary observables.

The reheating and CMB results for the three representative models in Figs.~\ref{fig:energy_density}--\ref{fig:eos_evolution} are
summarized in Table~\ref{tab:reheating_params}. For our baseline choice $\Gamma \simeq 2.435\times 10^{12}\,{\rm GeV}$, reheating completes within
$N_{\rm re}=\mathcal{O}(1)$ in all cases ($N_{\rm re}\simeq 0.57$--$1.55$), with reheating temperatures
$T_{\rm re}\sim(5.4\text{--}7.34)\times10^{14}\,\mathrm{GeV}$, which easily satisfy the BBN requirement
$T_{\rm re}\gtrsim \mathcal{O}({\rm MeV})$.
The mild spread in $(N_{\rm re},T_{\rm re})$ correlates with the two post-inflationary branches separated by the critical value $l_c$.
The $(3,10)$ benchmark lies in the $l<l_c$ branch, where the inflaton crosses the first barrier and reaches radiation domination more
quickly, giving a smaller $N_{\rm re}$ and a slightly higher $T_{\rm re}$. By contrast, $(1,10)$ and $(3,10^3)$ belong to the $l>l_c$
branch, in which the inflaton is trapped near the first minimum and oscillates, leading to a somewhat longer reheating stage and a
quasi-constant $\bar w(N)$ before it drifts toward $1/3$.
Using the matching condition~\eqref{eq:matching}, we obtain $N_*\simeq 55.40$--$56.37$ and the corresponding $(n_s,r)$ listed in
Table~\ref{tab:reheating_params}. While $n_s$ varies only mildly ($n_s\simeq 0.968$--$0.972$), $r$ shows a stronger model dependence,
ranging from $6.9\times10^{-4}$ to $1.8\times10^{-2}$. Overall, these predictions are compatible with the latest ACT DR6/Planck 2018
constraints and remain consistent with the BK18-driven upper limits on $r$ adopted in ACT+BK analyses.

\begin{table}[htbp]
\centering
\setlength{\tabcolsep}{12pt}      
\renewcommand{\arraystretch}{1.25}
\begin{tabular}{lccccc}
\hline
$(n,l)$ & $T_{\rm re}\,(\rm GeV)$ & $N_{\rm re}$ & $N_{\ast}$ & $n_s$ & $r$ \\
\hline
$(1,10)$     & $5.40\times 10^{14}$ & 1.55 & 56.37  & 0.972 & $1.8 \times10^{-2}$\\
$(3,10)$     & $7.34\times 10^{14}$ & 0.57 & 55.40  & 0.968 & $6.9\times10^{-4}$ \\
$(3,10^3)$   & $5.70\times 10^{14}$ & 1.33 & 56.02 & 0.969 & $6.5\times10^{-3}$ \\
\hline
\end{tabular}
\caption{Reheating and CMB parameters for $\Gamma=10^{-6}$ and the three representative parameter choices $(n,l)$ considered in this work.}
\label{tab:reheating_params}
\end{table}

\section{Conclusions}
\label{sec:conclusions}

We have studied single-field slow-roll inflation driven by the decaying oscillatory potential
$V(\phi)=\lambda\,\phi^{2n}\sin^2(l/\phi^n)$ and confronted its predictions with the latest ACT DR6 and Planck 2018 constraints,
using three representative benchmarks $(n,l)=(1,10)$, $(3,10)$, and $(3,10^3)$. Scanning $N_*$ generates trajectories in the
$(n_s,r)$ plane that can pass through the observationally allowed region. We find that increasing $n$ at fixed $l$ efficiently
suppresses $r$, whereas increasing $l$ at fixed $n$ tends to enhance $r$ with only a modest shift in $n_s$.

To go beyond the standard phenomenological reheating parametrization, we do not assume an ad hoc constant reheating equation-of-state parameter $w_{\rm re}$, nor scan over the usual reheating descriptors, namely the effective equation-of-state $w_{\rm re}$ during reheating and the reheating duration in $e$-folds $N_{\rm re}$, as well as the reheating temperature $T_{\rm re}$. Instead, we implement a minimal perturbative-decay reheating mechanism with a constant decay rate $\Gamma$ and numerically solve the coupled inflaton and radiation equations. This fully
dynamical treatment yields the time-dependent equation of state $w(N)$ and its cumulative average $\bar w(N)$, thereby determining
the macroscopic expansion history without extra assumptions.
As a concrete example, for our baseline $\Gamma\simeq 2.435\times 10^{12}\,{\rm GeV}$,
we find that reheating is efficient for all the representative parameter choices considered in this work: the system reaches
radiation domination within a small number of $e$-folds, $N_{\rm re}\simeq 0.5$--$1.5$, corresponding to a high reheating temperature
$T_{\rm re}\simeq (5$--$7)\times 10^{14}\,{\rm GeV}$ that comfortably satisfies the BBN requirement. Using the thermodynamic matching
relation, these reheating outcomes fix the pivot-scale $e$-fold number to a narrow range, $N_*\simeq 55$--$56$, and the associated
$(n_s,r)$ predictions fall well within the ACT-allowed region. Overall, incorporating a microphysically controlled perturbative
reheating history provides a more predictive link between the model parameters and ACT-era CMB constraints than the conventional
constant-$w_{\rm re}$ approach.

\begin{acknowledgments}

This work was supported by the National Natural Science Foundation of China (Grant No.~12505075), the Natural Science Foundation of Sichuan Province for Young Scientists, China (Grant No.~2026NSFSC0808), the Chengdu Normal University Talent Introduction Scientific Research Special Project (Grant No.~YJRC202443), the Research Incentive Program for Doctors Joining Shanxi (Grant No.~Z20240219), and the Fundamental Research Program of Shanxi Province (Grant No.~202303021212296).

\end{acknowledgments}

\bibliographystyle{apsrev4-2}
\bibliography{ref}

\end{document}